\documentclass[draft,jgrga]{AGUTeX}  %%%%ezaz

  \usepackage[dvips]{graphicx}
  \usepackage{xcolor}
\authorrunninghead{V\"{O}R\"{O}S ET AL.}

\titlerunninghead{Probability density functions}

\begin{document}

%% ------------------------------------------------------------------------ %%
%
%  TITLE
%
%% ------------------------------------------------------------------------ %%

%\title{Solar wind disturbances induced forced reconnection in the Earth's magnetosphere}
\title{Probability density functions for the variable solar wind near the solar cycle minimum}
%
% e.g., \title{Terrestrial ring current:
% Origin, formation, and decay $\alpha\beta\Gamma\Delta$}
% You may use \\ to break the title over several lines.

%% ------------------------------------------------------------------------ %%
%
%  AUTHORS AND AFFILIATIONS
%
%% ------------------------------------------------------------------------ %%

%Use \author{\altaffilmark{}} and \altaffiltext{}
\smallskip
\authors{Z. V\"or\"os, \altaffilmark{1, 6}
M. Leitner, \altaffilmark{2}
Y. Narita, \altaffilmark{1}
G. Consolini, \altaffilmark{3}
P. Kov\'acs, \altaffilmark{4}
A. T\'oth, \altaffilmark{5}
J. Lichtenberger, \altaffilmark{6}
 }

 \altaffiltext{1}{Space Research Institute, Austrian Academy of Sciences, Graz, Austria.}
 \altaffiltext{2}{University of Graz, Graz, Austria.}
\altaffiltext{3}{INAF-Istituto di Astrofisica e Planetologia Spaziali, Roma, Italy.}
\altaffiltext{4}{Geological and Geophysical Institute of Hungary, Budapest, Hungary.}
 \altaffiltext{5}{Constantine The Philosopher University, Nitra, Slovak Republic.}
  \altaffiltext{6}{Department of Geophysics and Space Sciences, E\"otv\"os University, Budapest, Hungary.}
%
% \altaffiltext{2}{FMI}
%
% \altaffiltext{3}{IFSI, Rome, Italy.}
%% ------------------------------------------------------------------------ %%
%
%  ABSTRACT
%
%% ------------------------------------------------------------------------ %%

% >> Do NOT include any \begin...\end commands within
% >> the body of the abstract.

\begin{abstract}
%\textcolor{red}{write words in red}
Unconditional and conditional statistics is used for studying the histograms of magnetic field multi-scale fluctuations in the solar wind near the solar cycle minimum in 2008.
The unconditional statistics involves the magnetic data during the whole year 2008. The conditional statistics involves the magnetic field time series splitted into concatenated subsets of data according to a threshold in dynamic pressure. The threshold separates fast stream leading edge compressional and trailing edge uncompressional fluctuations. The histograms obtained from these data sets are associated with both large-scale (B) and small-scale ($\delta$B) magnetic fluctuations, the latter corresponding to time-delayed differences. It is shown here that, by keeping flexibility but avoiding the unnecessary redundancy in modeling, the histograms can be effectively described by a limited set of theoretical probability distribution functions (PDFs), such as the normal, log-normal, kappa and log-kappa functions. In a statistical sense the model PDFs correspond to additive and multiplicative processes exhibiting correlations.
It is demonstrated here that the skewed small-scale histograms inherent in turbulent cascades are better described by the skewed log-kappa than by the symmetric kappa model. Nevertheless, the observed skewness is rather small, resulting in potential difficulties of estimation of the third-order moments. This paper also investigates the dependence of the statistical convergence of PDF model parameters, goodness of fit and skewness on the data sample size. It is shown that the minimum lengths of data intervals required for the robust estimation of parameters is scale, process and model dependent.
%The kappa functions are frequently used to describe the high-energy suprathermal tails of particle velocity distributions functions (VDFs) or skewed PDFs associated with turbulence or large-scale fluctuations in the solar wind. %While the suprathermal particle VDFs described by kappa functions can be explained in terms of the nonextensive statistical physics the large-scale skewed PDFs seem to be associated with a mixture of multiplicative processes %exhibiting correlations introduced by the solar ejecta. The small-scale magnetic fluctuations are associated with correlations in turbulence, nevertheless the underlying statistical features are also affected by the concomitant %large-scale structures, indicating the occurrence of cross-scale couplings.

\end{abstract}
%% ------------------------------------------------------------------------ %%
%
%  BEGIN ARTICLE
%
%% ------------------------------------------------------------------------ %%

% The body of the article must start with a \begin{article} command
%
% \end{article} must follow the references section, before the figures
%  and tables.

\begin{article}

%% ------------------------------------------------------------------------ %%
%
%  TEXT
%
%% ------------------------------------------------------------------------ %%

\section{Introduction}
The solar wind is a complex plasma system in which structures, wavy fluctuations and turbulence co-exist over multiple scales.
Since the Sun variably emits a mixture of fast and slow streams and transient ejecta, a single mechanism
cannot fully reproduce the observed complexity in the solar wind.
Some aspects of this complexity can be described statistically only.
Despite the structural and dynamical complexity, field and plasma fluctuations
exhibit well defined scalings over specific frequency ranges in power spectral density (PSD) log-log plots.
The low-frequency part of the spectrum is believed to reflect the statistics of large-scale structures and their interactions \citep{bur84}.
Below the timescale of a day the self-similar inertial range of magnetohydrodynamic (MHD) turbulence is set up, where dissipation
is supposed to be negligible and energy cascades towards the small scales (large wave numbers or high frequencies) \citep{bru13}.  At characteristic proton and
electron scales energy dissipation in collisionless plasmas takes place leading to non-adiabatic heating of the solar wind \citep{alex13}.

Turbulence represents a candidate process which explains the inertial range energy transfer supplying free energy for the kinetic scale dissipation processes.
The energy transfer rate $\epsilon$ in an MHD turbulent cascade can be calculated from the PSDs, however, these estimations are based on certain theoretical assumptions \citep{smith06, staw09}.
On the other hand, in hydrodynamics the rigorous Kolmogorov's 4/5 law
(Kolmogorov, 1941) represents a way to estimate $\epsilon$ from third-order structure functions. Similarly to the 4/5 law in hydrodynamics, a rigorous third-order moment relation to calculate $\epsilon$, the "MHD Yaglom's law",  was obtained for incompressible MHD case by \citet{poli98}.

In terms of probability distribution functions (PDFs) non-zero third order moments correspond to skewed PDFs. In hydrodynamic turbulence zero skewness or a non-skewed PDF indicates that the turbulent cascade is absent \citep{davi04}.
Therefore, the determination of $\epsilon$ should preferably be based on the estimation of third-order moments associated with skewed PDFs. However, direct evidence about the (non)stationarity of third-order moments, or equivalently, about the (non)stationarity of the asymmetric shape of PDFs  is still missing. In the solar wind the PDFs associated with turbulence are rather symmetric and it is difficult to estimate the skewness \citep{pod09}.

We will study the statistical properties of the large-scale fluctuations on the basis of total magnetic field measurements near the solar cycle minimum in 2008, when alternating fast and slow streams were present.
The small-scale statistical properties of two-point magnetic fluctuations will be evaluated on the basis of time-delayed data at the time scale of $\tau=$10 minutes. This is a time scale within the inertial range of turbulence in the solar wind. Parametric model PDFs will be used to fit the histograms obtained from the magnetic field data. The skewness (S) and the estimated parameters of model PDFs will be estimated separately for the large-scale and the small-scale magnetic fluctuations.

We note that the skewness S estimated from B is not straightforwardly associated with the turbulent energy transfer rate $\epsilon$. The third-order moment relation for calculating $\epsilon$ in the solar wind contains the components of the Elsässier field (combined magnetic and velocity fluctuations \citep{poli98}). Nevertheless, the MHD Yaglom's law, derived under the hypothesis of incompressibility, isotropy, homogeneity and stationarity, represents a rough approximation only, which is expected to describe certain aspects of MHD turbulence in the highly variable solar wind \citep{bru13}. Therefore, we believe, it makes sense to further investigate the statistical features of magnetic fluctuations alone, studying also the nonstationarity aspects of their dynamics.

According to \citep{bur04} the large-scale magnetic fluctuations are observed over time scales from hours to approximately one year. This means that the observation of the large-scale statistics is possible only by considering unfiltered long time series containing a representative number of the large-scale structures. High-pass filtering of the data, for example time delayed differencing, enhances the fluctuations over the small-scales. Although filtered long time series allow us to study the statistics of multi-scale fluctuations, the effect of large-scale structures on turbulence is unknown or studied only sporadically \citep{voro06}. We argue that conditional statistics applied for both unfiltered and filtered time series can reveal important aspects of multi-scale interactions.

We are going to show that the large-scale and the small-scale PDFs corresponding to the magnetic fluctuations during the whole year in 2008 are different from the conditional PDFs associated with magnetic fluctuations in the leading (compressional) and in the trailing edges of high speed streams. Nevertheless, a few basic types of model PDFs can explain the magnetic fluctuation statistics in the solar wind. These models correspond to additive and multiplicative statistical processes involving also nonzero correlations. We use four types of parametric PDFs, two symmetric distributions (normal and kappa) and two skewed distributions, log-normal and log-kappa. The basic mathematical features of normal and log-normal distributions were described long ago and their mathematical treatment can be find in textbooks. The kappa distribution arises from non-extensive statistical mechanics describing out of equilibrium systems with long-ranged correlations \citep{tsa88, leub02}. The log-kappa distribution was introduced recently by \citet{leit09}. As it is shown here, the meaning of the model PDF parameters and of the nonzero skewness depends on the time scale and on the nature of physical processes under consideration. As a consequence, the required sample size for the robust estimation of statistical parameters also varies according to the statistical conditioning.

The organization of the paper is as follows. Section 2 introduces the model PDFs and the methodology of the estimation of statistical parameters. In Section 3 unconditional and conditional statistics is introduced for both large-scale and small-scale magnetic fluctuations. The model PDF parameters and the skewness are estimated for progressively larger data sample sizes.
%This is interpreted then as a signature of higher-order statistical nonstationarity, which has to be considered in the following studies trying to estimate $\epsilon$.
Section 4 summarizes our results.
%and provides a short discussion of  possible

\section{Probability distribution functions (PDFs) for space plasmas}
We study magnetic field fluctuations over the time scales associated with large-scale structures and over the inertial range of scales associated with turbulent interactions. Our goal is to introduce a minimum number of model PDFs
which can explain substantial parts of magnetic field variability in the solar wind. To this end we will use the normal and kappa PDFs and their logarithmic counterparts.
%We will consider model PDFs arising from non-extensive statistical concepts.
The kappa PDFs which arise from non-extensive physics allow us to model magnetic field data exhibiting large-deviations from the normal distribution. Although kappa functions are commonly used to model non-Maxwellian particle velocity distribution functions (VDFs), the underlying physics associated with non-extensive statistics over the MHD and kinetic scales is rather different. To avoid confusion, we will shortly describe the differences between particle VDFs and magnetic field PDFs.

\subsection{Particle VDFs}
Particle VDFs in the solar wind exhibit low-energy near-Maxwellian core and high-energy suprathermal tails described by kappa functions \citep{vasil68, scud92, leub04}. Collisionless space plasmas are often out of thermal equilibrium and suprathermal particles can be accelerated by a number of mechanisms \citep{mar06, pier10, liva15}. There exists a strong connection between the long-tail kappa distributions or $\kappa$ indices (see below) and the non-extensive statistical mechanics representing a generalization of Boltzmann-Gibbs statistics for out of equilibrium particle systems exhibiting long-range correlations \citep{tsa88, leub02}. It has been shown the thermodynamic temperature $T$ can be uniquely defined for the non-extensive out-of-equilibrium systems as well, through the variance of VDFs (mean kinetic energy), or equivalently, through the connection of entropy with internal energy.
\citep{liva10}. The $\kappa$ index appears as an independent thermodynamic variable with a physical meaning of enhanced correlations (long-range interactions) or as a parameter corresponding to different stationary states in plasma systems with the same $T$ and $n$ (density) residing in a thermal non-equilibrium state \citep{liva13}. As a consequence, for the characterization of non-extensive plasmas described by kappa distributions three independent parameters are needed, $(n, T, \kappa)$. This can be regarded as a generalization of the polytropic law for out of equilibrium plasmas \citep{liva15}.

\subsection{PDFs for field and plasma statistics}
Examination of the field or particle statistics over the scales of MHD turbulence or even over larger scales implies that the real in-situ data contain contributions from different physical processes, structures or (non-)mixing plasmas with non-unique combinations of parameters $(n, T, \kappa)$. While the large-scale statistics can reflect important aspects of solar variability or locally generated processes in the solar wind, the variance of PDFs might not correspond to any well defined plasma temperature. For example, turbulence with the same types of plasmas might be confined to flux tubes having characteristic sizes within the inertial range of turbulence \citep{boro08}. Analyzing longer data intervals involving multiple flux tubes can change the turbulence statistics significantly \citep{boro08}. The frequently observed twisted flux tubes can be more unstable against Kelvin-Helmholtz instability or reconnection than unstable flux tubes \citep{zaqa14a, zaqa14b}. This can lead to abundant local generation of turbulence containing mixing plasmas and boundaries \citep{voro14}.

The length of the data sets is critical for reaching statistical convergence in parameter estimations. In fact, careful examination of the radial velocity and radial magnetic field statistics
in the solar wind has shown that
%the moments are close to zero (the PDFs are rather symmetric) and
$10^6$-$10^7$ data points are  required for robust estimation of the third-order moments associated with skewed PDFs \citep{pod09}. In other studies of third-order moment estimations, based on the Els\"assier variables, the number of data points was a few thousands only \citep{sor07}. However, it was argued by \citet {sor10} that turbulence in the fast and slow streams has different properties, therefore merging data into longer data sets for reaching statistical convergence can be misleading.
%In a statistical study \citet{bur04} have already analyzed the PDFs corresponding to magnetic field and speed fluctuations during the declining phase of the solar cycle 23 in 2003. They considered 275 days in their statistical %study, which included multiple corotating streams, corotating interaction regions and turbulence.
PDFs corresponding to magnetic field and speed fluctuations during the declining phase of the solar cycle 23 in 2003 were analyzed by \citet{bur04}. They considered 275 days in their statistical study, which included multiple corotating streams, corotating interaction regions, discontinuities and turbulence. Nevertheless, the authors found that the multi-scale fluctuations can be well fitted by the non-extensive model PDFs \citep{bur04}.

As one can see from the above studies, long data intervals are needed for the robust estimation of statistical parameters associated with multi-scale structures and turbulence. On the other hand, the inclusion of structures, boundaries, solar ejecta, compressional regions and turbulence, occurring during long time intervals, would most certainly results in non-unique combinations of  $(n, T, \kappa)$. This would preclude the connection of model kappa PDFs with non-extensive statistics
or particle VDFs. There is a clear need for a better understanding of the meaning of kappa PDFs over the larger scales which contain mixtures of different plasmas.

\subsection{Parametric PDFs for the magnetic field}
We consider four types of parametric distribution functions which correspond to additive, multiplicative, non-extensive statistical processes and their combinations.
%It is not yet clear, however, what is their role in understanding turbulence statistics in space plasmas.

The symmetric Gaussian (normal) distribution $P_n$ appears as a result of the additive central limit theorem: the sum of identically distributed independent random variables (each having an expected value $\mu$ and variance $\sigma ^2$) is normally distributed:
\begin{equation}
  P_n(x, \mu, \sigma) = \frac{1}{\sigma \sqrt{ \pi}} \ \ exp\left[ {-\frac{(x-\mu)^2}{\sigma^2}} \right],
  \ \ \ \ \ \ \ -\infty < x < \infty,
 \end{equation}
It has already been found that PDFs corresponding to the time delayed differences $\delta (X(t,\tau))=X(t+\tau )-X(t)$, calculated from solar wind time series $X(t)$, do not follow the normal distribution $P_n$ over the small time scales $\tau$  \citep{mar97, sor99, bur04}. This is the time scale of turbulence, coherent structures, interaction regions, spanning from hours down to seconds. Due to correlations and nonlinear interactions the small-scale PDFs are peaked, skewed and exhibit fat tails (the largest fluctuations occur more frequently than for normally distributed random fluctuations). However, random fluctuations over longer time scales $\tau$ (from hours to days) exhibit Gaussian distribution due to the loss of correlations.

The skewed log-normal distribution $P_{Ln}$ appears as a result of multiplicative central limit theorem \citep{crow88} and reads as:
\begin{equation}
  P_{Ln}(x, \mu, \sigma) = \frac{1}{x} \ \ \frac{1}{\sigma \sqrt{\pi}} \ \ exp\left[-\frac{(\log x - \mu)^2}{\sigma^2} \right],
    \ \ \ \ \ \ \ 0 < x < \infty,
\end{equation}
Here $\mu$ is a scale parameter while $\sigma$ represents a shape parameter.

When no time delayed differentiation is applied, the magnetic field, density, proton temperature statistics exhibit skewed PDFs which can be approximately modeled  through the log-normal distribution \citep{bur00, brun04, ves10, dmi13}. Since log-normal distributions are generated through multiplications of random variables, logarithmic scale invariance in turbulence is often understood in terms of multiplicative cascades \citep{cas90}. In the solar wind random multiplications beyond turbulence are also foreseen for random amplification/weakening of waves, or for compressional effects increasing the values of proton density or magnetic field \citep{dmi13}. In fact, the log-normal model describes rather well the density and temperature fluctuations with the highest probabilities (within $\pm 2 \sigma$) or the magnetic field statistics (roughly within $\pm 1.5 \sigma$). However, the fat tails present in data distributions show again probabilities higher than the $P_{Ln}$ model \citep{dmi13}. This is also valid for the histograms of derived quantities, e.g. for the histogram of the solar wind quasi-invariant (the inverse square of the Alfv\'en-Mach number), which correlates well with sunspot cycle \citep{leit11a}. We also mention that instead of a single log-normal PDF two or more different log-normal models are added to form a single skewed one to fit the fluctuation statistics associated with the jumps in magnetic field vector orientation and  magnetic field intensity \citep{brun04}. It is supposed that different $P_{Ln}$ models might be associated with different components of MHD turbulence, such as  uncompressive fluctuations of Alfv\'enic origin or compressive structures. Three component log-normal models can also be associated with the anisotropic magnetic field-aligned (slab) and field-perpendicular (quasi-2D) populations of fluctuations complemented by convected compressive structures \citep{brun04}. It is not known how many different log-normal PDFs were needed to model the solar wind data.

A superposition of random uncorrelated, normally or lognormally distributed processes can provide a composed skewed PDF with fat tails which can fit the observations reasonably well. However, turbulent space plasmas contain structures, long-range interactions and correlations. Normal (or for particles the Maxwell-Boltzmann) distributions are associated with Boltzmann-Gibbs statistics describing systems with non-interacting or weakly interacting subsystems in thermal equilibrium. For stationary non-equilibrium systems the Boltzmann-Gibbs entropy was generalized and the so-called nonextensive entropy was introduced by \citet{tsa98}. The nonextensive entropy introduces interactions between subsystems (e.g. particles) \citep{leub02} and its extremization under certain physically relevant constraints leads to kappa distribution \citep{tsa98, liva13}. A simple physical model based on the generalized Galton board clearly demonstrates the formation of fat tailed kappa-like distributions from Gaussian distributions, when interaction terms and memory effects are switched on \citep{leit11b}. The peaked, fat tailed data distributions corresponding to turbulence in the solar wind are well modeled by the kappa distribution \citep{leub05a, leub05b}:
\begin{equation}
 P_{\kappa}(x, \mu, \sigma, \kappa) =   \frac{1}{ \sigma \sqrt{\kappa \pi} } \ \  \frac{\Gamma (\kappa)}{\Gamma (\kappa-1/2)} \ \ \left(1+  \frac{(x-\mu)^2}{\sigma^2 \kappa}   \right)^{-\kappa},
   \ \ \ \ \ \ \ -\infty < x < \infty,
\end{equation}
The nonextensive $\kappa$ index represents a measure of correlations or memory in systems out of equilibrium, when $\kappa \ll \infty $.
%the peaked and fat tailed $\kappa $ distribution is ...

Since the kappa distribution  represents a symmetric distribution, the observed skewed distributions \citep{bur00} or the turbulent energy transfer rate $\epsilon$ \citep{pod09} cannot be modeled by $P_{\kappa}$.
\citet{bur04} added a somewhat artificial cubic term to their version of $P_{\kappa}$ which could be used then to fit the observed skewed PDFs. An alternative approach to deal with the skewed PDFs was formally based on the assumption that the $\kappa$-distributed correlated fluctuations can also be combined in a multiplicative manner leading to the log-kappa distributions \citep{leit09}:
\begin{equation}
 P_{L\kappa}(x, \mu, \sigma, \kappa) =   \frac{1}{x} \ \ \frac{1}{\sigma \sqrt{\kappa \pi} } \ \  \frac{\Gamma (\kappa)}{\Gamma (\kappa-1/2)} \ \ \left(1+  \frac{(\log x-\mu)^2}{\sigma^2 \kappa}   \right)^{-\kappa}
   \ \ \ \ \ \ \ 0 < x < \infty,
\end{equation}

The normal, log-normal, kappa and log-kappa distributions are compared to each other in Figure 1. The nice feature of these model PDFs is that for $\kappa \rightarrow \infty $, $P_{\kappa} \Rightarrow P_n$ (Figure 1a) and $P_{L\kappa} \Rightarrow P_{Ln}$ (Figure 1b). In other words, depending on the values of $\kappa$ (we show  $P_{\kappa}$ and  $P_{L\kappa}$ for $\kappa=$3, 10, 20), the kappa and log-kappa distributions are capable to reproduce the PDF tails which are fatter than the normal or log-normal PDFs.
Although other types of parametric PDFs (e.g. \citep{con09}) or their combinations could also work well, we provide evidence that the considered PDFs are sufficient to model substantial part of the observed magnetic fluctuations in the variable solar wind in 2008.

\subsection{Fitting and error statistics}
The Levenberg-Marquardt nonlinear least-square method \citep{press92} will be used to fit the parametric model PDFs to the histograms of the large-scale and time-delayed magnetic field fluctuations. The algorithm initially uses the gradient descent method to iteratively update the fitting parameters in large steps, then it adaptively switches to Gauss-Newton method minimizing the sum of squares in smaller steps. Together with the fitted curves the 95-percent
confidence intervals of the fits will be plotted. The error measure for the parameters is estimated as the squared-root of the diagonal of the parameter covariance matrix. The goodness of fit $Q$ is calculated from the incomplete gamma function $gammq$ as $Q = gammq(0.5\nu, 0.5\chi ^2)$, where $\nu $ is the number of degrees of freedom and $\chi ^2$ is the "chi-square" parameter calculated from minimizing the sum of squares divided by the standard deviation \citep{press92}. The dependence of the model PDF parameters on the length of the time series will be calculated as well. Additionally, the skewness $S$ (describing the asymmetry of PDFs) and its standard error will be estimated using a bootstrap method. The bootstrap is a Monte-Carlo simulation method treating the original time series as a pseudo-population from which new populations are obtained via random resampling \citep{marti07}.

\section{Data analysis}
In this paper we use OMNI 1 minute high resolution plasma and magnetic data (\citep{king05} and references therein) near the solar cycle minimum in 2008. The subplots in Figure 2 a-c show the bulk speed (V), the magnetic field magnitude (B) and the dynamic pressure ($P_{dyn}$). The alternating fast speed streams are interacting with the slower plasma leading to compressions in front of the streams \citep{bur04} and enhanced values of B and $P_{dyn}$. The trailing edge of the fast streams is characterized by lower values of B and  $P_{dyn}$. The threshold  $P_{dyn}=$3 [nPa], indicated by the red dashed horizontal line in Figure 2c roughly separates the compression dominated leading edges from the trailing edges of the streams. Figure 3 shows the scatterplots of B versus $P_{dyn}$, including two different streams (blue and black points for each) from the beginning of 2008. Again, the red dashed line corresponding to the threshold $P_{dyn}=$3 [nPa] separates the compressional leading edge from the trailing edge fluctuations. Since the compressions can produce lognormally distributed multiplications of the magnetic field \citep{dmi13} which are not present in the trailing edge regions of the streams, we may consider the possibility that the corresponding magnetic fluctuations are ruled by different statistical laws. To test this we consider the magnetic fluctuation statistics for the whole year 2008 and conditional statistics for the subsets of the data defined by B($P_{dyn}>3)$ [nPa] and by B($P_{dyn}<3)$ [nPa]. For the yearly statistics we consider the whole data set in 2008. For conditional statistics data corresponding to the short sudden transitions at the front of the leading edges of streams are discarded. Mixed data intervals with alternating short duration large amplitude pressure fluctuations, containing intervals with $P_{dyn}$ below and over the 3 [nT] threshold, are also omitted. The compressional intervals represent roughly 5\% and the trailing edge intervals 50\% of the 527040 data points in 2008.

\subsection{Large-scale PDFs}

Figures 4 a1, b1 and c1 show the histograms corresponding to total magnetic field fluctuations B (blue points) and the parametric model PDF fits ($P_{Ln}$, $P_{L\kappa}$ and $P_{n}$) with confidence intervals. The maxima of the distributions are normalized to 1. Figures 4 a2, b2 and c2 show the absolute values of residuals $R$, corresponding to the difference between data and PDF models. The histogram for the whole year 2008 is shown in Figure 4 a1. The fits and the black and red residuals in Figures 4 a1, a2, corresponding to $P_{Ln}$ and $P_{L\kappa}$, respectively,  indicate that the magnetic fluctuations near the maximum of PDF are described by both distributions equally well. This agrees well with the results of \citet{dmi13} that the fluctuations with the highest probabilities near the PDF maxima are log-normally distributed. Our analysis demonstrates, however, that the fat tails of the histogram are much better described by the log-kappa model. These results describe fluctuation statistics over long time intervals associated with multiple physical processes and mixtures of plasmas.

Let us consider now conditional statistics for the subsets of compressional (B($P_{dyn}>3)$ [nPa]) and trailing edge (B($P_{dyn}<3)$ [nPa]) data. These cases are shown in Figures 4 b1, b2 and 4 c1, c2, respectively. As one can see, the compressional magnetic fluctuations form a skewed distribution while the trailing edge fluctuations are symmetric and normally distributed. The tails of the histogram associated with compressional fluctuations are described slightly better by the the log-kappa PDF than the log-normal PDF. From the conditional PDFs (Figures 4 b1, c1) it becomes clear how the unconditional PDF (Figure 4 a1) associated with the yearly mixture of processes is formed. The normally distributed trailing edge data encompassing the fluctuations between 0 and 8 [nT] (Figure 4 c1) make the main contribution to the central part of the yearly PDF (Figure 4 a1).
The rather good $P_n$ model fit indicates that the normally distributed independent uncompressive magnetic fluctuations within the different trailing edges additively form a final normal distribution.
Within the range of 0 - 20 [nT], the skewed parts of the yearly PDF or the fat tails (Figure 4 a1) are formed by the compressional magnetic fluctuations (Figure 4 b1). The $P_{L\kappa}$ model provides a better fit of the tails than the $P_{Ln}$ model indicating the compressive fluctuations are associated with multiplicative processes which additionally involve long-range correlations.

The dependence of parameters ($\kappa, \sigma, \mu$) for the $P_{L\kappa}$ and $P_n$ models on the length of the time series is shown in Figures 5a-c. The horizontal axis indicates the increasing length of the time series (sample size) for which the PDF model parameters and their errors are estimated with the above described Levenberg-Marquardt algorithm. The color code corresponds to the unconditional data (blue), compressional (black) and trailing edge data (red). The data sets have different maximal lengths. The parameters are estimated for data intervals of increasing lengths until the maximal lengths are reached. In Figure 5d the goodness of fit parameter $Q$ is shown. The PDF models are statistically rejected when $Q \rightarrow 0$, for example $Q \ll 0.05$.

The PDF models are satisfactory when $Q > 0.05$ and the fitted parameters show stationarity. According to these criteria the unconditional statistics calculated from the yearly data (blue curves) is well represented by the $P_{L\kappa}$ model when the lengths of data intervals are longer than 30-50 days, for which $Q > 0.05$. For the same intervals $\kappa \sim 2$ (blue curve in Figure 5a), indicating strong deviation of the log-kappa model from the log-normal model (Figure 1b). Roughly the same amount of trailing edge data (red curves in Figure 5)  is required to reach statistical significance with the $P_n$ model. However, when more than 40 day long intervals are considered, the parameters $\sigma$ and $\mu$ show a quasi-stationary behavior. Since the $P_n$ model is associated with a conditional statistics, the required length of data intervals are longer than 40 days in the real time. The lengths of data intervals for the robust estimation of unconditional $P_{L\kappa}$ and trailing edge $P_{n}$ statistics indicate that the underlying magnetic fluctuations might be associated with successive 27-day quasi-periodic recurrent streams. This is also valid for the large-scale compressional statistics (black curves in Figure 5). After more than 10 days of data $Q$ becomes larger than 0.05 (Figure 5d). However, due to the B($P_{dyn}>3)$ [nPa] condition, the compressional data are collected from multiple streams during the whole year. Since the log-kappa and log-normal PDFs are rather similar (Figure 4 b1) the $\kappa$ parameter is large. At $Q > 0.05$ (Figure 5d) $\kappa = 8 \pm 4$ (Figure 5a). The relatively large errors ($\pm 4$) are caused by the small amount of compressional data ($\sim$ 24000 data points).

The skewness ($S$) associated with the conditional and unconditional statistics is depicted in Figure 5d. The same color code is used as above. The error bars calculated from bootstrap are negligible. In accordance with the skewed PDFs (Figures 4 a1 and 4 b1), both the yearly and compressional data show positive skewness $S \sim 2$, while the skewness associated with normal PDF (Figure 4 c1) is zero (Figure 5d).

The results of the model PDF fits together with $S$ are in Table 1. The values of parameters correspond to the full lengths of data sets.
\subsection{Small-scale PDFs}
Figure 6 shows the histograms, PDF model fits and the absolute residuals for the time delayed differences $\delta$B = B(t+$\tau$)-B(t), where $\tau=$10 minutes. The same notation is used as in Figure 4.
Since the histograms are rather symmetric, the $P_{\kappa}$ and $P_{n}$ model fits are compared. The residuals show (Figures 6 a2-c2) that both models describe the central parts of the PDFs equally well. However, the kappa PDFs are better suited for the modeling of the fat tails. Despite the normally distributed large-scale fluctuations (Figure 4 c1) the small-scale trailing edge PDF is peaked and long-tailed (Figure 6 c1). At the time scale of  $\tau=$10 minutes these fluctuations can be associated with inertial range turbulence. The same is true for the compressional fluctuations in Figure 6 b1. However, the distribution is wider and large deviations have higher probabilities than in the trailing edge case. The unconditional yearly PDF in Figure 6 a1 obviously contains a mixture of sudden jumps, trailing edge and compressional fluctuations, all with different physical properties.

The dependence of parameters ($\kappa, \sigma, \mu$) for the $P_{\kappa}$ model on the length of the time series is shown in Figures 7a-c. The goodness of fit parameter $Q$ and the skewness $S$ are shown in Figures 7d and 7e, respectively. The small-scale $\kappa $ indices vary between 1.5 and 2 indicating strong correlations and fat tails at the given time scale $\tau$. The parameter $Q$ shows that the required length for the robust estimation of the model parameters is below 1 day for the conditional and 7-10 days for the yearly unconditional statistics. In comparison with the large-scale fluctuations, shorter data intervals are required for the robust estimation of model parameters over the small scales. This indicates, that for the reliable estimation of the skewness, or perhaps the turbulent energy transfer rate $\epsilon$,  several hours or few days long data intervals might suffice.
The results of the model PDF fits together with $S$ are in Table 2. The values of parameters correspond to the full lengths of data sets.

A nonzero $\epsilon$ should be associated with skewed distributions, however, our PDFs are rather symmetric (Figure 6). \citet{pod09} have pointed out that the third order moments or the skewness represent signed moments subject to cancelation effects. Indeed, the skewness $S$ calculated for the different $\delta$B data change sign in Figure 7e. Also, the error bars are of the same order as the mean values indicating the occurrence of strong non-stationarity effects. By increasing the sample size the cancelation of successive positive and negative skewnesses can effectively lead to $S\sim 0$ hamstringing the calculation of the actual value of $\epsilon$. In fact, $S$ approaches zero for the largest sample sizes in Figure 7e. In hydrodynamics a skewness of zero would indicate no vortex stretching and no energy cascade in a turbulent flow \citep{davi04}. Positive or negative $S$ could be associated with typical coherent structures such as sheets or tubes generated by vortex stretching or compression \citep{davi04}. In our case $S$ simply changes sign when $\delta$B = -B(t+$\tau$)+B(t) is considered rather than $\delta$B = B(t+$\tau$)-B(t). Therefore, one-point magnetic measurements of time delayed differences do not allow the identification of sheet or tube like structures. When only the absolute values of $S$ are considered separately for the trailing edge events then the yearly average PDF becomes skewed at the given scale $\tau$ with $S=0.25 \pm 0.1$. %(indicated by asterisk in Table 2).

A larger $S$ can be obtained by considering subsets of the data.
As an example, we selected three subsets of trailing edge data in 2008, each with $S\sim -0.38$ at $\tau = 10$ minutes. Bootstrap calculations for the concatenated data, consisting of 25000 data points, give $S=-0.38 \pm 0.08$. Figure 8 shows the histogram of time delayed magnetic fluctuations shifted by 4 [nT] to positive values. The shift does not affect the skewness, however, it allows a comparison of symmetric and skewed PDF models. Figure 8a shows the $P_{\kappa}$ (symmetric, black line) and $P_{L\kappa}$ (asymmetric, red line) model fits. The absolute residuals in Figure 8b show that the central part of the histogram is fitted by the $P_{\kappa}$ and $P_{L\kappa}$ models equally well. The tails of the histogram are described slightly better by the skewed $P_{L\kappa}$. Nevertheless, the fluctuations with the smallest probabilities deviate from the PDF models, possibly indicating finite size effects. Naturally, there exist several factors which can affect the statistical convergence of statistical parameters, such as $S$, $\epsilon$ or the parameters of model PDFs. Our results show that beside the sample size the identification and separation of underlying physical processes are equally important.
%Quite contrarily,
%$S$ exhibits negligible errors for the total magnetic field fluctuations in Figure 5e.

\section{Conclusions}
The main goal of this paper was to show that a limited number of parametric model PDFs can describe major parts of the magnetic fluctuations over multiple scales in the solar wind. The range of scales from minutes to several months predetermines the nature of physical processes occurring near a solar cycle minimum in the solar wind.

The large scales are dominated by a mixture of corotating streams, corotating interaction regions, solar ejecta, discontinuities and turbulence \citep{bur04}. The large-scale statistics is based on the histograms of the total magnetic field fluctuations B obtained in 2008.
This paper shows that the histogram associated with the large-scale mixture of physical processes  is best described by the log-kappa $P_{L\kappa}$ model (Figure 4). Naturally, the $\kappa$ parameter has a completely different meaning for the large-scale PDFs than for the almost instantaneous particle VDFs. Although the non-extensive interpretation of the $\kappa$ index for the large-scale magnetic histograms is not straightforward, the log-kappa PDF model seems to be superior to the log-normal model even for longer time intervals encompassing a solar cycle \citep{leit11a}. This indicates that the $\kappa$ index could have a reasonable interpretation over the large-scales as well. In order to further decipher the physics behind the large-scale log-kappa model we used conditional PDFs, separating magnetic fluctuations according to the dynamic pressure. Conditional probability in space physics content has already been advocated by \citet{ukhor04} to describe the multi-scale features in solar wind-magnetosphere coupling. In our case, the pressure criteria separated fluctuations in space and time, roughly disengaging fluctuations at the leading edge of the fast streams where $P_{dyn}>$3 [nPa], from the fluctuations at the trailing edge of the fast streams where $P_{dyn}<$3 [nPa].
It was demonstrated that the fluctuations of the magnetic field within the concatenated trailing edges are Gaussian distributed. It is not unexpected, the time delayed differences $\delta (B(t,\tau))=B(t+\tau )-B(t)$ for large
$\tau$ also exhibit a normal distribution \citep{sor99, bur04}. Since for large enough $\tau$ the correlations in turbulence are lost, the large-scale trailing edge PDF appears to be the signature of uncorrelated magnetic fluctuations which are not affected by the small-scale turbulent cascades. Contrarily, the concatenated leading edge magnetic fluctuations associated with $P_{dyn}>$3 [nPa] are non-symmetric. In this case the $\kappa $ parameter in the skewed $P_{L\kappa}$ model is large ($\kappa=8\pm 4$, Table 1) and the log-kappa and log-normal PDF models are very close to each other. The "logarithmic" models associated with compressive fluctuations indicate the occurrence of multiplicative processes which has already been found by \citet{ves10} and \citet{dmi13}. However, when the yearly PDF is considered, the mixture of normally distributed trailing edge and the lognormally distributed leading edge processes clearly forms a log-kappa distribution with $\kappa=3.5\pm 0.5$ (Figure 4), providing a better fit than the $P_{Ln}$ model. The other two parameters associated with $P_{Ln}$ and $P_{L\kappa}$ models ($\mu$ and $\sigma$ in Table 1) are similar. At this stage, however, it is impossible to interpret the $\kappa$ index in terms of the non-extensive physics. The mixture of large-scale processes cannot be associated with a combination of (n,T,$\kappa$) parameters. Also, it is neither possible nor relevant to interpret the skewness associated with yearly or compressional large scale fluctuations ($S=2.4\pm0.1$ and $S=2\pm0.1 $, Table 1) as a quantity which could be related to energy transfer rate in a turbulent cascade. Nevertheless, we offer here the interpretation that the large-scale $\kappa$ index varies as the occurrence frequency of multiplicative processes, that is, the occurrence of stream and ejecta interaction regions changes from year to year. This can influence the  average properties of the solar wind, such as "the average amount of turbulence" or "the average rate of plasma heating" across a solar cycle. It has to be checked by further careful examination of the large-scale statistics.  At the present time we have no models which could describe the large-scale variations of average properties of the solar wind plasma containing correlated and uncorrelated structures and a mixture of fluctuations.

The small-scale statistics is based on the histograms of the time-delayed differences $\delta (B(t,\tau))=B(t+\tau )-B(t)$ obtained in 2008. The chosen time delay is $\tau= 10$ minutes thus the small scales are dominated by a mixture of turbulent fluctuations. The histograms at this scale in Figure 6 are best described by the kappa distributions with $\kappa$ between 1.1 and 1.6 (Table 2). According to the non-extensive approach the kappa values are restricted to the interval $\kappa \in (1.5, \infty ]$, simply because for $\kappa < 1.5$ the temperature is not determined for kappa VDFs \citep{pier10, liva13}. The limiting value $\kappa = 1.5$ corresponds to the furthest state from equilibrium (or "anti-equilibrium") while $\kappa \rightarrow \infty$ corresponds to the equilibrium state associated with a Maxwellian VDF \citep{liva13}. It would be tempting to interpret the estimated values $\kappa \sim 1.5$ for $\delta B$ PDFs as a signature of a thermodynamic anti-equilibrium state. However, the histograms obtained from $\delta B$ involve the yearly and concatenated mixtures of data and a unique anti-equilibrium state would not account for the multiple physical processes occurring in 2008. We believe it is safer to say that the low values of $\kappa$ index at the scale $\tau$ are associated with the structures repeatedly generated by turbulence introducing correlations to observed fluctuations. We mention that correlations in a generalized Galton board lead to peaked and fat tailed kappa-like distributions, while in the absence of correlations Gaussian distributions are obtained \citep{leit11b}.
Figure 6 and the fitting results in Table 2 show that the $\kappa$ and $\sigma$ parameters are significantly different for the yearly and for each conditional PDFs. Although each histogram is peaked and fat tailed, the leading edge compressional data seem to be associated with different type of turbulence than the trailing edge data. This might indicate that the large-scale structures which are selected by the conditional statistics are affecting the small-scale turbulence.

The goodness of fit parameter $Q$ depends on the data sample size. It is not surprising that for the robust estimation of large-scale model PDF parameters longer data sets are needed than for the small-scale model PDF parameters
(Figures 5 and 7). The proper sample size ensuring a good fit, however, does not guarantee that the skewness $S$ can be reliably calculated. $S$ is a signed moment therefore subject to cancelation effects. The consideration of longer data sets of $\delta$B containing alternating positive and negative $S$ leads to $S\rightarrow 0$ which can be seen in Figure 7e. When only those subsets of data are considered for which the skewness is, for example, equal and negative, the skewed $P_{L\kappa}$ model describes the tails of the histogram slightly better than the symmetric $P_{\kappa}$ model.

Finally, we mention that for the description of normal or peaked, fat tailed or skewed histograms the log-kappa and kappa PDF models might suffice. From these models the normal and log-normal distributions can be obtained by taking the limit $\kappa \rightarrow \infty$. Although other types of model PDFs or their combinations could also describe the peaked and skewed histograms well (\citep{con09}), the kappa and logarithmic kappa models allow to study the multi-scale changes of histograms in a simple way. Anyhow, for a systematic study of the statistical variability in the solar wind, model PDFs have to be chosen. The kappa family of model PDFs seems to offer the desirable flexibility avoiding the unnecessary redundancy in modeling. Further studies are needed, however, to establish a firm connection between the non-extensive physics and the kappa distributions associated with turbulent or large-scale fluctuations. In order to estimate properly the skewness and the energy transfer rate, it would also be important to extend this analysis by considering conditional statistics for other physical quantities, including the Els\"assier variables.

\begin{figure}
\noindent\includegraphics[width=32pc]{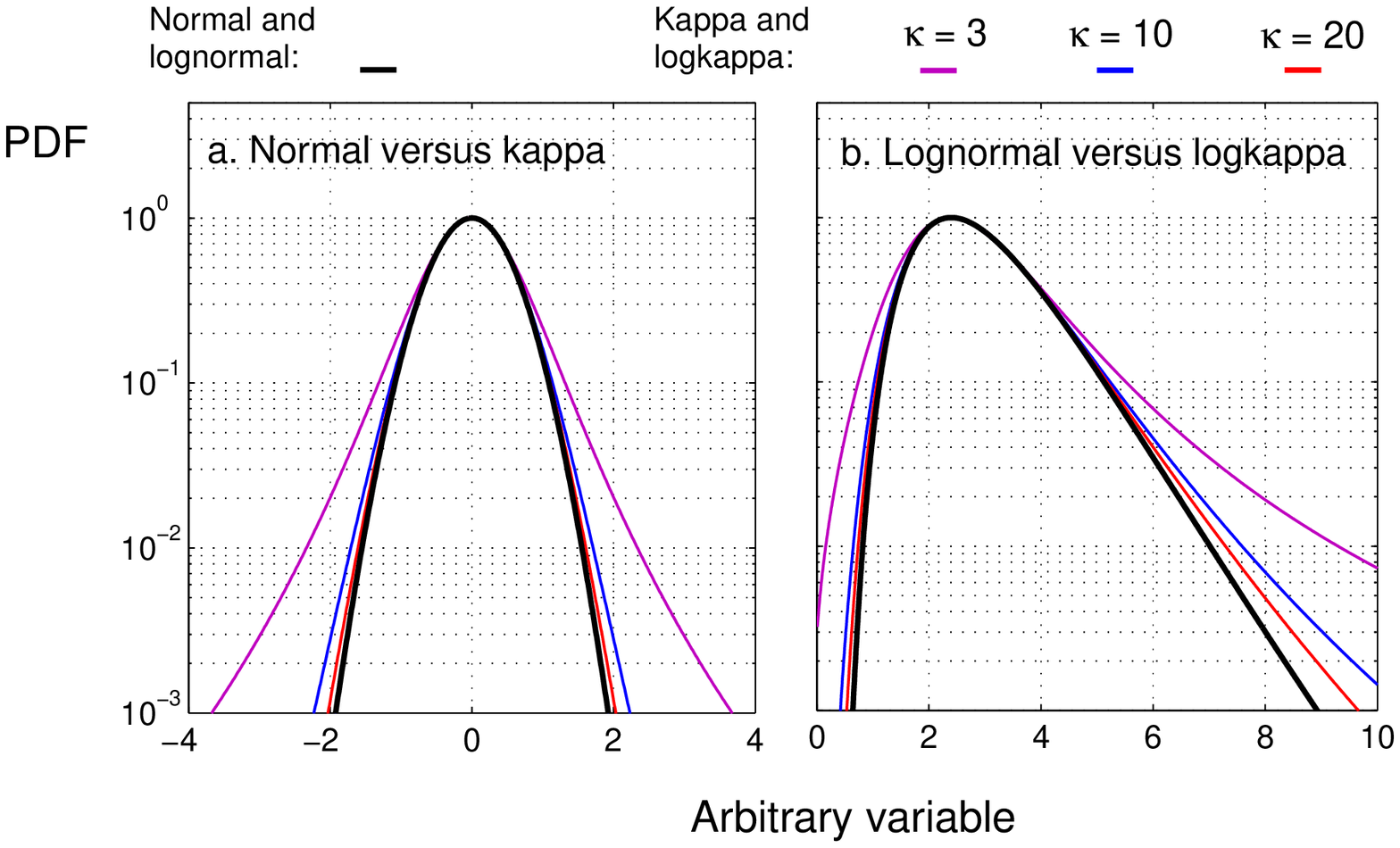}
 \caption{Comparison of theoretical probability density functions (PDFs) for different $\kappa$ parameters.
 a.) Normal versus kappa model PDFs; b.) Log-normal versus log-kappa model PDFs.) }
 \end{figure}

\begin{figure}
\noindent\includegraphics[width=32pc]{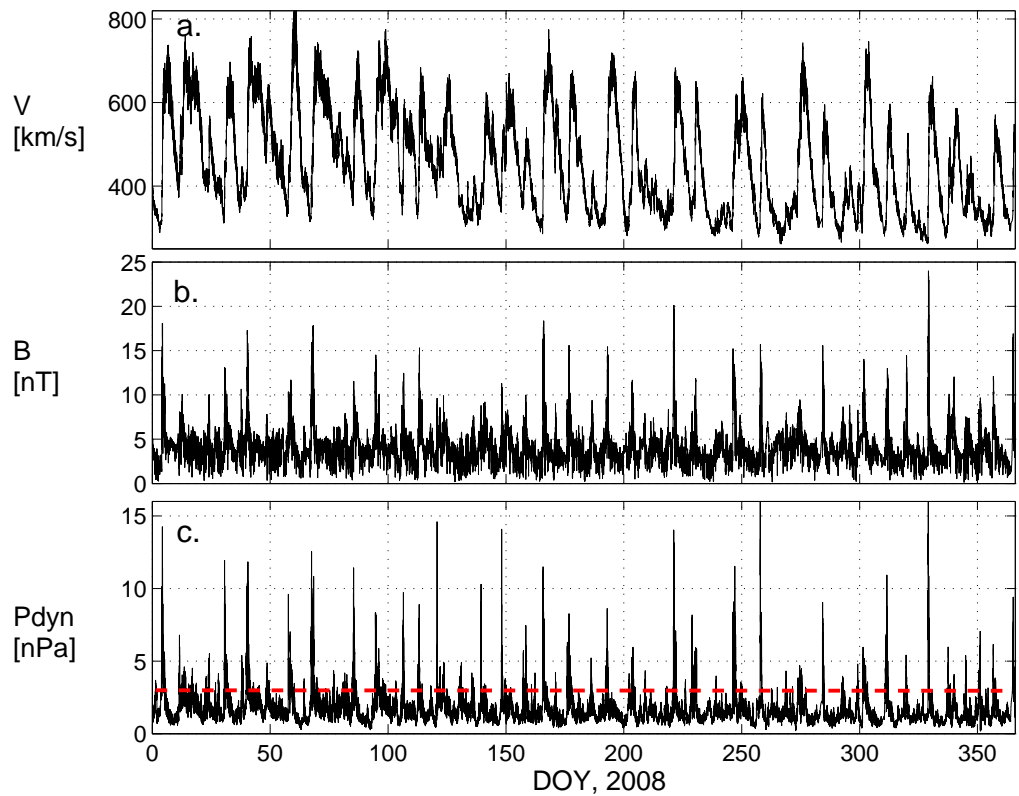}
 \caption{OMNI solar wind data near the solar cycle minimum in 2008. a.) Bulk speed V; b.) Total magnetic field B; c.) Dynamical pressure $P_{dyn}$.
 The red dashed horizontal line indicates the threshold separating compressional data ($P_{dyn}>$3 nPa) at the leading edge of high speed streams from the uncompressional data
 ($P_{dyn}<$3 nPa) at the trailing edge of high speed streams.}
 \end{figure}

\begin{figure}
\noindent\includegraphics[width=32pc]{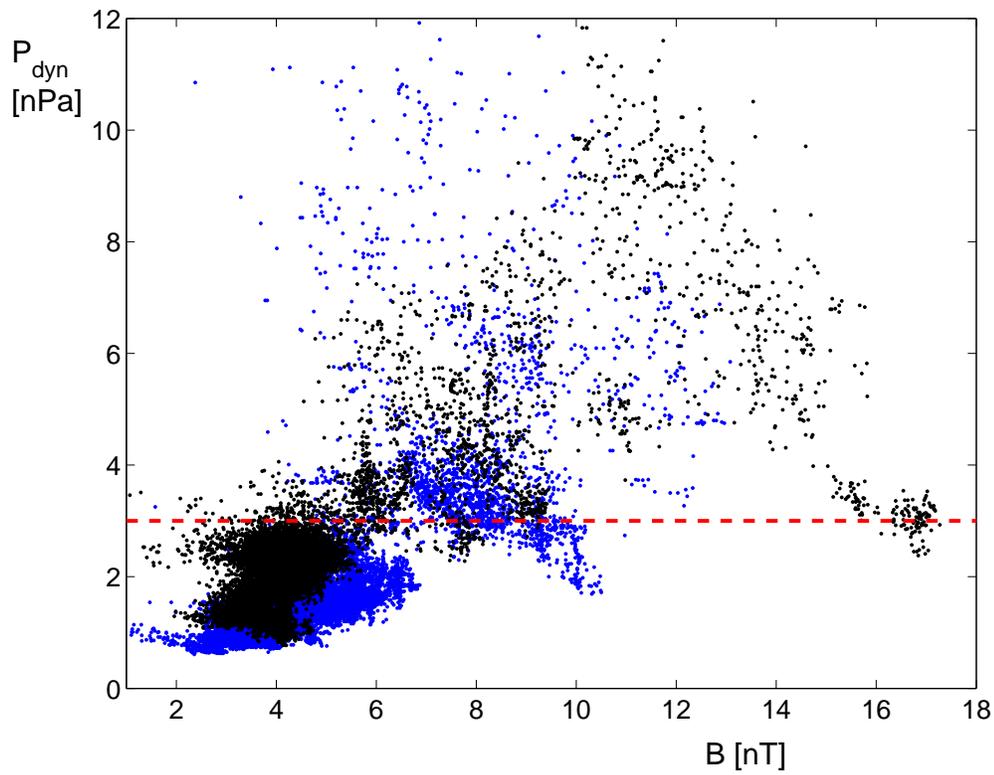}
 \caption{Scatterplot of magnetic field magnitude B versus dynamic pressure $P_{dyn}$. Two different high-speed streams are shown (blue and black points) from the beginning of 2008. The threshold $P_{dyn}$=3 nPa is indicated by the red dashed line.}
 \end{figure}

\begin{figure}
\noindent\includegraphics[width=32pc]{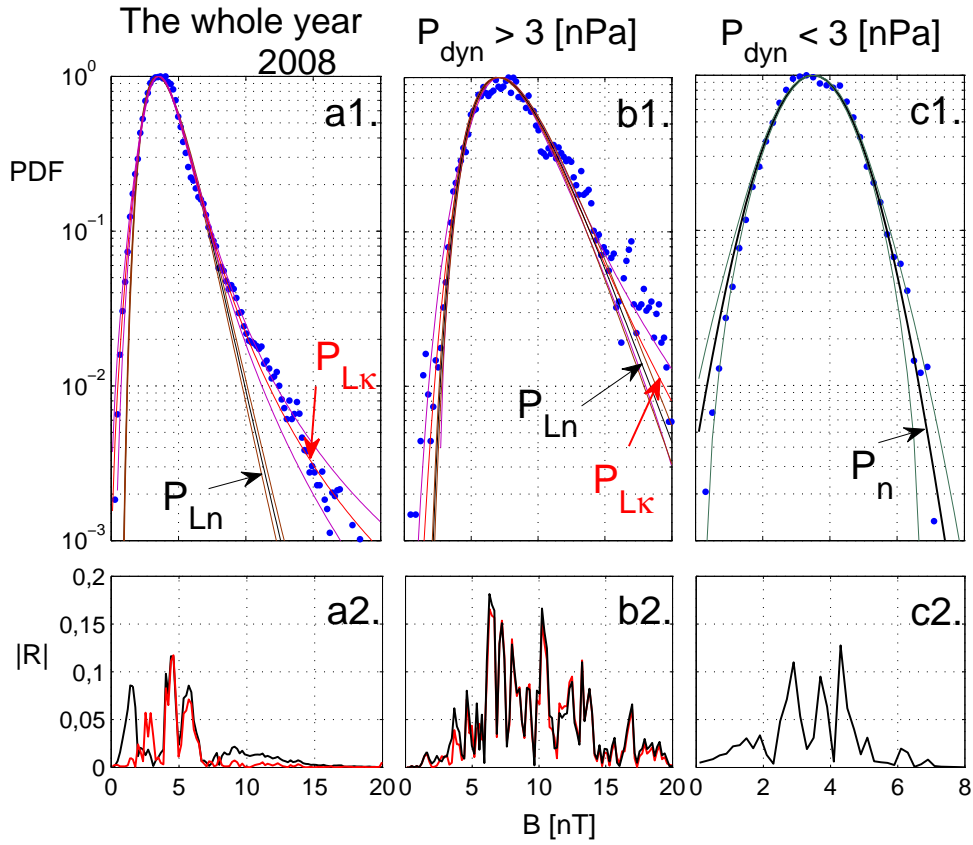}
 \caption{Comparison of the histograms of the total magnetic field fluctuations (B) with model PDFs and the absolute values of the residuals $|R|$. a1.)-a2.) Statistics for the whole year in 2008; b1.)-b2.) Conditional statistics of the concatenated leading edge compressional intervals; c1.)-c2.) Conditional statistics of the concatenated trailing edge uncompressional intervals. Normal ($P_n$), log-normal ($P_{Ln}$) and log-kappa ($P_{L\kappa}$) model PDFs are shown with 95\% confidence intervals.}
 \end{figure}

\begin{figure}
\noindent\includegraphics[width=39pc]{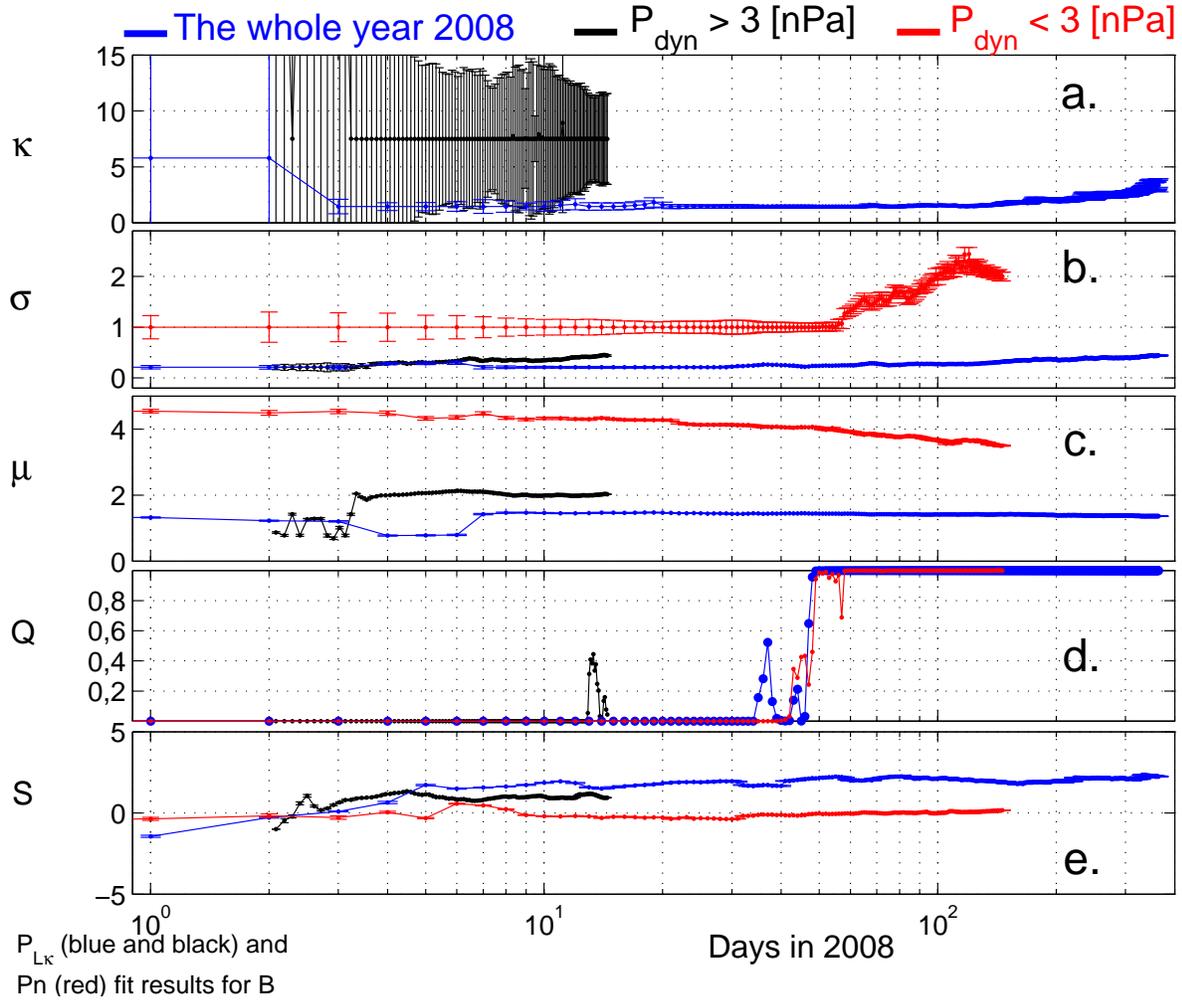}
 \caption{Dependence of fitting parameters on the data sample size for unconditional (blue) and conditional (black and red) total magnetic field (B) statistics, including the error bars. a.)-b.)-c.) model PDF parameters $\kappa$, $\sigma$, $\mu$; d.) goodness of fit parameter Q; e.) skewness S; The color code for conditional and unconditional statistics is indicated on the top of the figure; besides, the red color also corresponds to the $P_n$, the red and blue colors to the $P_{L\kappa}$ model fits. $\sigma$, $\mu$ are in [nT] for $P_n$ and dimensionless for $P_{L\kappa}$.
 }
 \end{figure}

\begin{figure*}
\noindent\includegraphics[width=39pc]{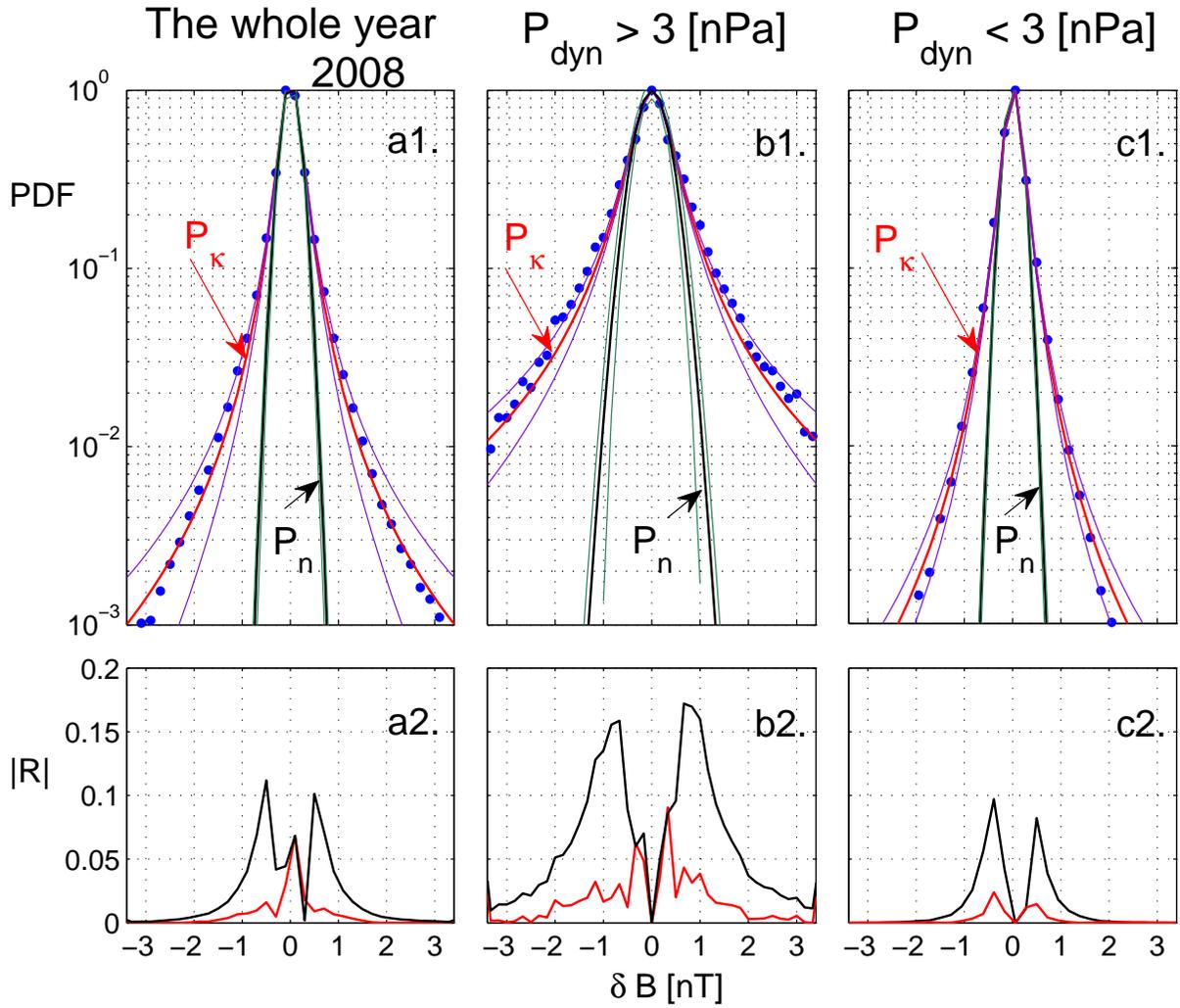}
 \caption{Comparison of the histograms of the time delayed magnetic field fluctuations ($\delta$B) with model PDFs and the absolute values of the residuals $|R|$. a1.)-a2.) Statistics for the whole year in 2008; b1.)-b2.) Conditional statistics of the concatenated leading edge compressional intervals; c1.)-c2.) Conditional statistics of the concatenated trailing edge uncompressional intervals. Normal ($P_n$) and kappa ($P_{\kappa}$) model PDFs are shown with 95\% confidence intervals. }
 \end{figure*}

 \begin{figure*}
\noindent\includegraphics[width=39pc]{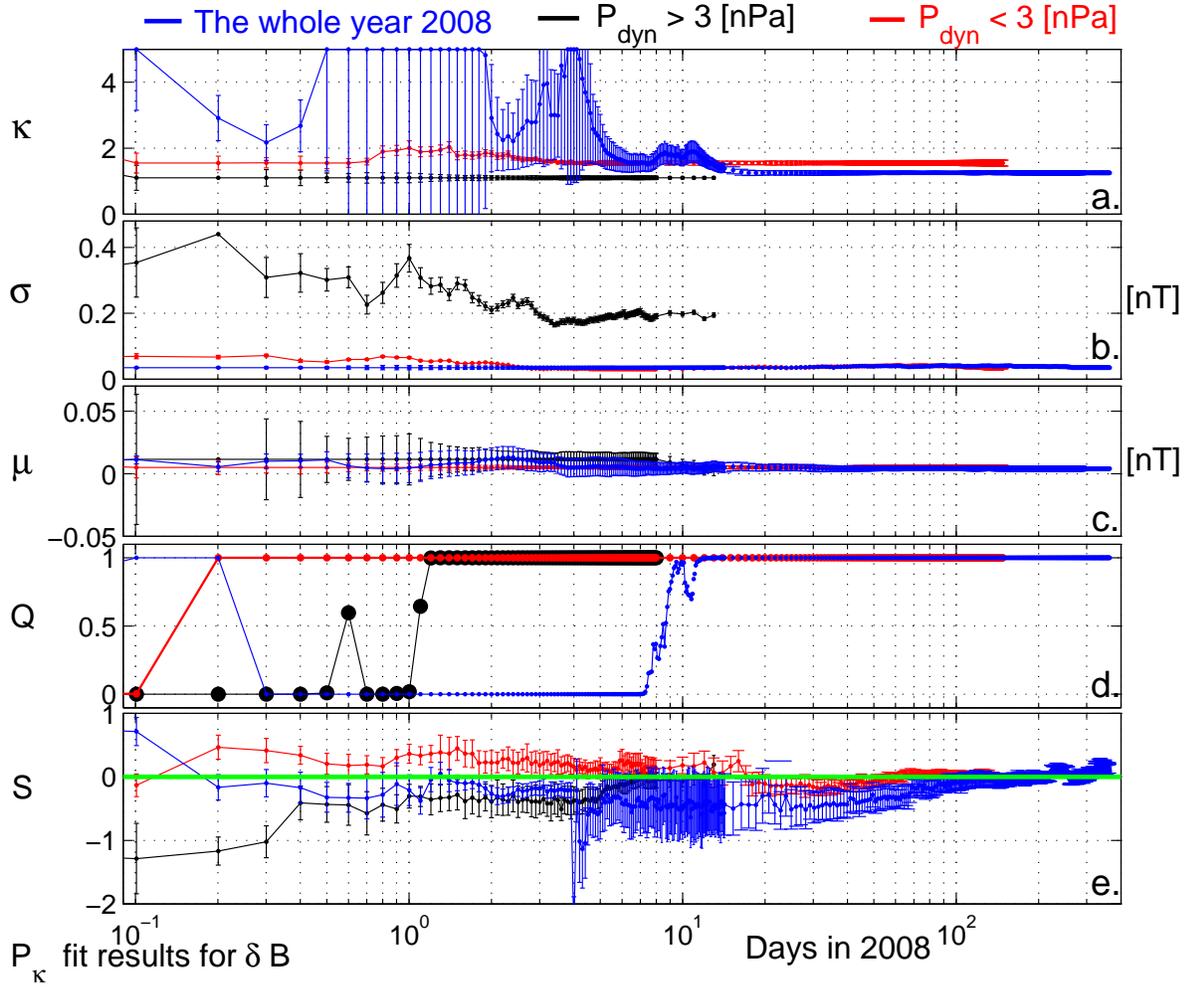}
 \caption{Dependence of fitting parameters on the data sample size for unconditional (blue) and conditional (black and red) time delayed magnetic field ($\delta$B) statistics with error bars. a.)-b.)-c.) $P_{\kappa}$ model parameters $\kappa$, $\sigma$, $\mu$; d.) goodness of fit parameter Q; e.) skewness S; The color code for conditional and unconditional statistics is indicated on the top of the figure.
 }
 \end{figure*}

\begin{figure}
\noindent\includegraphics[width=39pc]{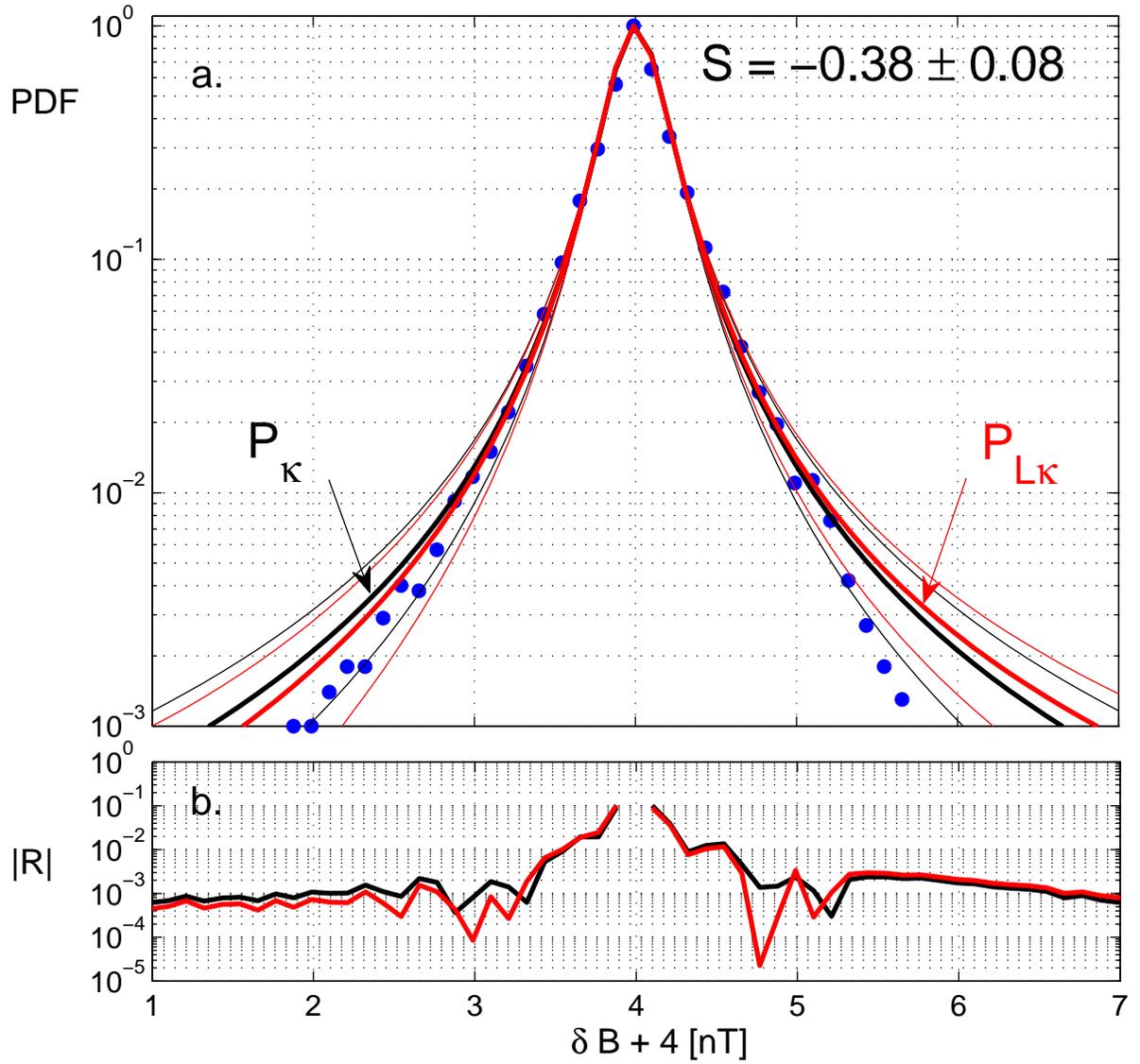}
 \caption{Comparison of the histograms of the time delayed magnetic field fluctuations ($\delta$B + 4 nT) with model PDFs and the absolute values of the residuals $|R|$. Three subsets of trailing edge data in 2008 are concatenated, each with $S\sim -0.38$. The asymmetric $P_{L\kappa}$ model represents a slightly better fit of the fat tails than the symmetric $P_{\kappa}$ model.}
 \end{figure}

\begin{table}
\begin{tabular}{|l|c|c|c|c|c|}
  \hline
  Time series & Skewness & PDF & $\mu$ & $\sigma$ & $kappa$ \\
  \hline
  \hline
  B                   & 2.4$\pm$0.1 &  $P_{Ln}$      & 1.36$\pm$0.01 & 0.49$\pm$0.01 &  \\
                      &             &  $P_{L\kappa}$ & 1.36$\pm$0.01 & 0.42$\pm$0.05 & 3.5$\pm$0.5   \\
  \hline
  B ($P_{dyn}>3$ nPa) & 2$\pm$0.1 &  $P_{Ln}$      & 2.06$\pm$0.01 & 0.46$\pm$0.01 &  \\
                      &             & $P_{L\kappa}$  & 2.06$\pm$0.01 & 0.44$\pm$0.01 & 8$\pm$4 \\
  \hline
  B ($P_{dyn}<3$ nPa) & 0.1$\pm$0.1 & $P_{n}$        & 3.5$\pm$0.01   & 2$\pm$0.15    &  \\
\hline
\end{tabular}
\caption{PDF model fitting statistics for the large-scale magnetic fluctuations (B). $\sigma$, $\mu$ are in [nT] for the $P_n$ model.}
\end{table}

\begin{table}
\begin{tabular}{|l|c|c|c|c|c|}
  \hline
  Time series & Skewness & PDF & $\mu$ & $\sigma$ & $kappa$ \\
  \hline
  \hline
  $\delta$B                    & 0.15$\pm$0.1  &  $P_{n}$      &   0 & 0.032$\pm$0.01   &  \\
                               &                 &  $P_{\kappa}$ &   0 & 0.03$\pm$0.01    & 1.2$\pm$0.05   \\
  \hline
  $\delta$B ($P_{dyn}>3$ nPa)  & 0.1$\pm$0.1 &  $P_{n}$      & 0   & 0.2$\pm$0.05     &  \\
                               &                 & $P_{\kappa}$  & 0   & 0.18$\pm$0.05    & 1.1$\pm$0.2 \\
  \hline
  $\delta$B ($P_{dyn}<3$ nPa)  & 0$\pm$0.05    & $P_{n}$       & 0   & 0.04$\pm$0.002    &  \\
                               & 0$\pm$0.05    & $P_{\kappa}$  & 0   & 0.039$\pm$0.002   & 1.58$\pm$0.05 \\
                               %& * -0.38$\pm$0.08    & $P_{\kappa}$  & 0   & 0.039$\pm$0.002   & 1.58$\pm$0.05 \\
                               %& * -0.38$\pm$0.08    & $P_{L\kappa}$  & 0   & 0.039$\pm$0.002   & 1.58$\pm$0.05 \\
\hline
\end{tabular}
\caption{PDF model fitting statistics for the time-delayed magnetic fluctuations ($\delta$B). $\sigma$, $\mu$ are in [nT] for the $P_n$ and the $P_{\kappa}$ models. }
\end{table}

%%% End of body of article:

%%%%%%%%%%%%%%%%%%%%%%%%%%%%%%%%
%% Optional Appendix goes here
%
%%%%%%%%%%%%%%%%%
% Geophysical Research Letters only allows an appendix without a letter.
%% You can get this result with
%  \section*{Appendix}
%  or
%  \section*{Appendix: Title}
%%%%%%%%%%%%%%%%%
%
% \appendix resets counters and redefines section heads
% but doesn't print anything.
% After typing  \appendix
%
% \section{Here Is Appendix Title}
% will print
% Appendix A: Here Is Appendix Title
%
% \section*{Appendix}
% will print
% Appendix
%
% \section*{Appendix: Here Is Appendix Title}
% will print
% Appendix: Here Is Appendix Title
%
% For only 1 appendix \appendix \section{Appendix} is preferred.
% which will print
% Appendix A

%%%%%%%%%%%%%%%%%%%%%%%%%%%%%%%%%%%%%%%%%%%%%%%%%%%%%%%%%%%%%%%%
%
% Optional Glossary or Notation section, goes here
%
%%%%%%%%%%%%%%
% Glossary only allowed in Reviews of Geophysics
% \section*{Glossary}
% \paragraph{Term}
% Term Definition here
%
%%%%%%%%%%%%%%
% Notation -- End each entry with a period.
% \begin{notation}
% Term & definition.\\
% Second Term & second definition.
% \end{notation}
%%%%%%%%%%%%%%%%%%%%%%%%%%%%%%%%%%%%%%%%%%%%%%%%%%%%%%%%%%%%%%%%
%
%  ACKNOWLEDGMENTS

\begin{acknowledgments}
This work was supported by the Austrian Fond zur F\"{o}rderung der wissenschaftlichen Forschung (projects P24740-N27)
and by  EU collaborative project STORM - 313038.
The OMNI data were obtained from the GSFC/SPDF OMNIWeb interface at http://omniweb.gsfc.nasa.gov.
\end{acknowledgments}

%% ------------------------------------------------------------------------ %%
%
%  REFERENCE LIST AND TEXT CITATIONS
%
% Either type in your references using
% \begin{thebibliography}{}
% Text
% \end{thebibliography}
%
% Or,
%
% If you use BiBTeX for your References, please produce your .bbl
% file and copy the contents into your paper here.
%
% Follow these steps:
% 1. Run LaTeX on your LaTeX file.
%
% 2. Run BiBTeX on your LaTeX file.
%
% 3. Open the new .bbl file containing the reference list and
%   copy all the contents into your LaTeX file here.
%
% 4. Comment out the old \bibliographystyle and \bibliography commands.
%
% 5. Run LaTeX on your new file before submitting.
%
% AGU does not want a .bib or a .bbl file, but asks that you
% copy in the contents of your .bbl file here.

%Reference citation examples:

%...as shown by \textit{Kilby} [2008].
%...has been shown [\textit{Kilby et al.}, 2008].

%...as shown by \cite{jskilby}.
%...has been shown \citep{jskilbye}.

%% ------------------------------------------------------------------------ %%
%
%  END ARTICLE
%
%% ------------------------------------------------------------------------ %%

\end{article}

%% Enter Figures and Tables here:

% When submitting articles through the GEMS system:
% COMMENT OUT ANY COMMANDS THAT INCLUDE GRAPHICS.

% Figure captions go below this illustration; Table captions go above tables

% ONE-COLUMN figure/table, including eps graphics
%
% \begin{figure}
% \noindent\includegraphics[width=20pc]{samplefigure.eps}
% \caption{Caption text here}
% \end{figure}
% \end{document}
%
% \begin{table}
% \caption{}
% \end{table}
%
% ---------------
% TWO-COLUMN figure/table
%
% \begin{figure*}
% \noindent\includegraphics[width=39pc]{samplefigure.eps}
% \caption{Caption text here}
% \end{figure*}
%
% \begin{table*}
% \caption{Caption text here}
% \end{table*}
%
% see below for how to make landscape figures or tables

%%% End the article here:

\end{document}